%% file: ArXiv_Version16052021.tex
\documentclass[
superscriptaddress,
frontmatterverbose, 
preprint,
showpacs,preprintnumbers,
 amsmath,amssymb,
 natcommun,
floatfix,
]{revtex4-1}
\usepackage{appendix}
\usepackage{SIunits}
\usepackage{sistyle}
\usepackage[utf8x]{inputenc}
\usepackage{graphicx}   
\graphicspath{{fig/}}   
\usepackage{dcolumn}
\usepackage{bm}
\usepackage{hyperref}
\setcitestyle{super}
\usepackage{epstopdf}
\usepackage{lipsum}
\usepackage{braket}
\DeclareGraphicsExtensions{.pdf,.png,.jpg,.jpeg,.mps}
\usepackage{pgf}
\usepackage{tikz}
\usepackage{xr}
\usepackage{standalone}
\usepackage{changes}
\usepackage{blindtext}
\usepackage[symbol]{footmisc}

\begin{document}
\input{Manuscript_MAC_revised_cleancopy_arxiv.tex}

\newpage

\input{SupportingInformation_revised_cleancopy_arxiv.tex}
\end{document}

%% file: Manuscript_MAC_revised_cleancopy_arxiv.tex
\title {Voltage-Controlled Reconfigurable Magnonic Crystal at the Submicron Scale}

\author{Hugo Merbouche}
\altaffiliation{These authors contributed equally}
\affiliation{Unit\'{e} Mixte de Physique CNRS, Thales,  Universit\'{e}  Paris-Saclay, 91767 Palaiseau, France}

\author
{Isabella Boventer$^{*}$}

\email{Corresponding author: isabella.boventer@cnrs-thales.fr}

\affiliation{Unit\'{e} Mixte de Physique CNRS, Thales,  Universit\'{e}  Paris-Saclay, 91767 Palaiseau, France}

\author{Victor Haspot}

\altaffiliation{These authors contributed equally}
\affiliation{Unit\'{e} Mixte de Physique CNRS, Thales,  Universit\'{e}  Paris-Saclay, 91767 Palaiseau, France}


\author{Stéphane Fusil}
\affiliation{Unit\'{e} Mixte de Physique CNRS, Thales,  Universit\'{e}  Paris-Saclay, 91767 Palaiseau, France}

\address{Universit\'{e} d’Evry, Universit\'{e} Paris-Saclay, 91000 Evry, France}
\author{Vincent Garcia}
\affiliation{Unit\'{e} Mixte de Physique CNRS, Thales,  Universit\'{e}  Paris-Saclay, 91767 Palaiseau, France}

\author{Diane Gou\'{e}r\'{e}}
\affiliation{Unit\'{e} Mixte de Physique CNRS, Thales,  Universit\'{e}  Paris-Saclay, 91767 Palaiseau, France}

\author{Cécile Carr\'{e}t\'{e}ro}
\affiliation{Unit\'{e} Mixte de Physique CNRS, Thales,  Universit\'{e}  Paris-Saclay, 91767 Palaiseau, France}

\author{Aymeric Vecchiola}
\affiliation{Unit\'{e} Mixte de Physique CNRS, Thales,  Universit\'{e}  Paris-Saclay, 91767 Palaiseau, France}

\author{Romain Lebrun}
\affiliation{Unit\'{e} Mixte de Physique CNRS, Thales,  Universit\'{e}  Paris-Saclay, 91767 Palaiseau, France}

\author{Paolo Bortolotti}
\affiliation{Unit\'{e} Mixte de Physique CNRS, Thales,  Universit\'{e}  Paris-Saclay, 91767 Palaiseau, France}

\author{Laurent Vila}

\address{Universit\'{e} Grenoble Alpes, CEA, CNRS, Grenoble INP, Spintec, 38000 Grenoble, France}
\author{Manuel Bibes}
\affiliation{Unit\'{e} Mixte de Physique CNRS, Thales,  Universit\'{e}  Paris-Saclay, 91767 Palaiseau, France}

\author{Agnès Barth\'{e}l\'{e}my}
\affiliation{Unit\'{e} Mixte de Physique CNRS, Thales,  Universit\'{e}  Paris-Saclay, 91767 Palaiseau, France}

\author{Abdelmadjid Anane}
\email{Corresponding author: madjid.anane@universite-paris-saclay.fr}
\affiliation{Unit\'{e} Mixte de Physique CNRS, Thales,  Universit\'{e}  Paris-Saclay, 91767 Palaiseau, France}
\begin{abstract}
Multiferroics offer an elegant means to implement voltage-control and on the fly reconfigurability in microscopic, nanoscaled systems based on ferromagnetic materials.  These properties are particularly interesting for the field of magnonics, where spin waves are used to perform advanced logical or analogue functions. Recently, the emergence of nano-magnonics {\color{black} is expected to} eventually lead to the large-scale integration of magnonic devices. However, a compact voltage-controlled, on demand reconfigurable magnonic system has yet to be shown. Here, we introduce the combination of multiferroics with ferromagnets in a fully epitaxial heterostructure to achieve such voltage-controlled and reconfigurable magnonic systems. Imprinting a remnant electrical polarization in thin multiferroic $\mathrm{BiFeO_3}$  with a periodicity of $500\,\mathrm{nm}$ yields a modulation of the effective magnetic field in the micron-scale, ferromagnetic $\mathrm{La_{2/3}Sr_{1/3}MnO_3}$ magnonic waveguide. We evidence the magneto-electrical coupling by characterizing the spin wave propagation spectrum in this artificial, voltage induced, magnonic crystal and demonstrate the occurrence of a robust magnonic bandgap with $>20 \,\mathrm{dB}$ rejection.
%
\end{abstract}
\maketitle
Spin waves, the collective spin excitations in a magnetically ordered material and their associated quanta, magnons, are promising candidates to be used as information carriers for the next generation of information processing devices \cite{Lenk2011,Kruglyak2010,Khitun2010}. Among all beyond-CMOS (complementary metal-oxide semiconductor) based approaches, magnonics combines room-temperature operation at GHz frequencies, nano-scale integration, digital and analog architectures \cite{Kostylev2005,Schneider2008a,Zografos2017,Ustinov2010}. Magnons can interact directly or indirectly with frequencies from the microwave to the optical regime. Thus, magnonics enables the coupling to other paradigms such as quantum computing and opto-electronics rendering it a rich playground to test innovative ideas to overcome the power consumption bottleneck of conventional CMOS-based electronics \cite{Tabuchi2015,Lachance-Quirion2019,Lachance-Quirion2020,Kusminskiy2016,Parvini2020, Chumak2015, Chumak2019}. However, the compatibility with state of the art CMOS based technologies will still be required for any practical magnonic device. Therefore, signal transducing and processing should be advantageously performed through voltage rather than current driven signals. Correspondingly, piezo-electric transducers and modulators have been proposed \cite{Duflou2017,Sadovnikov2018}. In wave-based devices, signal processing most often relies on artificial crystals. Magnonic artificial crystals (MACs) have therefore been extensively studied \cite{Chumak2017,Qin2018,Frey2020,Choudhury2020,Merbouche2021}. Achieving reconfigurability has however proven to be difficult. Indeed, most practical realisations of MACs are based on a periodic structuration which prohibits any reconfiguration of the frequency response. To date, reconfigurability has only been demonstrated in a few cases relying on the magnetic stray field modulation, albeit the spatial period of the MAC is fixed by construction \cite{Chumak2009a,Topp2010, Krawczyk2014}.  
Recently, theoretical propositions based on micromagnetic simulations have shown the interest of voltage-controlled reconfigurable MAC \cite{Wang2017}. Such devices based on the switching of multiferroic domain states would simultaneously allow for electric field control and low-power operation \cite{Hu2016,Rana2019}.
\\ Here, we present a fully reconfigurable, voltage-controlled device which passively couples magnonics to the magneto-electric degrees of freedom in a multiferroic/ferromagnetic heterostructure (\figref{Fig1:ExpSetup}) and operates at $300\,\mathrm{K}$. 
Regarding the multiferroic, we employ a $23\,\mathrm{nm}$ thick layer of  $\mathrm{BiFeO}_3$ (BFO). This material is one of the most studied multiferroics and exhibits both a sizeable magneto-electric coupling and a piezo-electric effect of $40 \,\mathrm{pm/V}$ \cite{Ramesh2007,Sando2014,Park2014,Garcia2015,Spaldin2017,Spaldin2019}. 
For the ferromagnetic layer which serves as the spin wave medium, we utilize a $13 \,\mathrm{nm}$ thick layer of $\mathrm{La}_{2/3}\mathrm{Sr}_{1/3}\mathrm{MnO}_3$ (LSMO) \cite{Hemberger2002}. 
Based on this structure, we create our reconfigurable and tuneable crystal by using the piezo-force microscopy (PFM) technique to generate a uniformly or periodically polarized ferroelectric pattern in the BFO \cite{Alexe2004,Gruverman2019}. In turn that pattern modulates the spin wave propagation in the LSMO, and allows to obtain spin wave filtering at the magnonic bandgap of the crystal. 
We identify the creation of the bandgap \textemdash the hallmark of a magnonic crystal \textemdash by a careful comparison of the transmitted spin wave spectra whilst the adjacent ferroelectric domain state is changed between a uniform down (virgin) or up or periodically down-up ferroelectric polarization. 
That comparison (cf. \figref{Fig.2SpinWavePropagationLSMO}) allows us to unambiguously attribute the observed frequency filtering (cf. \figref{Fig3GapIndication}) in the amplitude representation at $3.54\,\mathrm{GHz}$ to a bandgap in the magnonic medium due to the periodic domain configuration of BFO. {\color{black} In our study we show the modulation of the spin wave propagation by applying a voltage to the BFO layer which switches the ferroelectric polarization states. Our conceptual realization of such voltage control allows for {\color{black}other} perspectives to go beyond PFM based writing.  For instance, the theoretical propositions for  all-solid state devices discussed by the work of  Q. Wang {\it{et al.}} \cite{Wang2017} can be adapted to the BFO polarisation control using an array of voltage-gates. Writing or erasing the MAC could be simply performed using voltage spikes on a desired set of electrodes. }
\\\\\
\section*{Results and Discussion}
\begin{figure}
\centering
\includegraphics[scale=0.35]{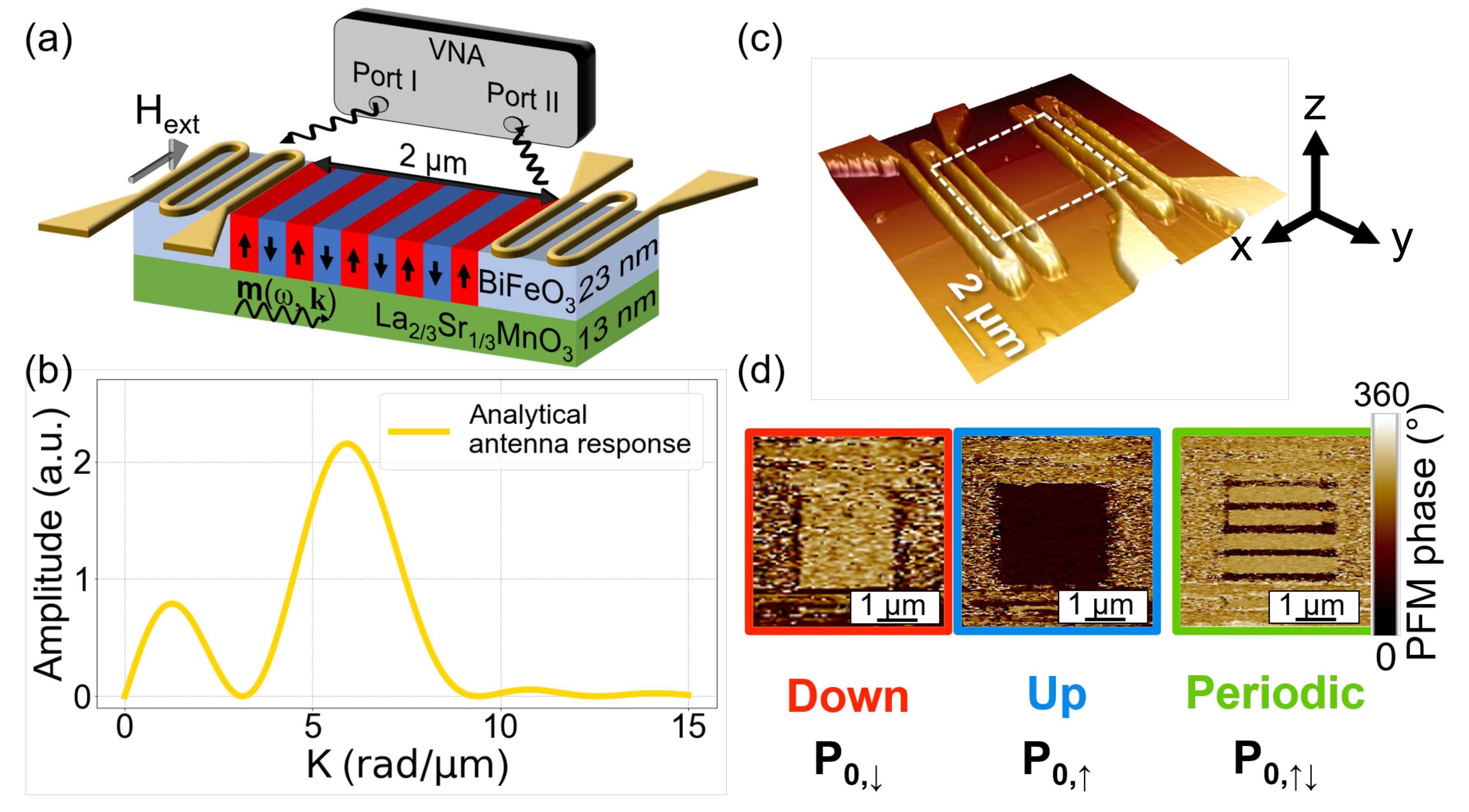}
\caption{(a) Experimental setup of our BFO/LSMO heterostructure  shown for the periodic configuration to obtain a reconfigurable magnonic crystal on the sub-micron scale. The  spin wave is excited in Damon-Eshbach configuration and the ferroelectric domain state is written \textit{via} piezo-force microscopy (PFM). The inductance matrix for a two-port system using vector network analysis is measured \textit{via} probe-tip ferromagnetic resonance {\color{black} at room temperature ($T=300\,\mathrm{K}$)} (b) Analytical calculation of the antenna's response function with maximum at wavevector  $K \approx 6\, \mathrm{rad/\mu m}$. 
(c) Image of the radio frequency antennae with spacing of $500\,\mathrm{nm}$ and $2\,\mathrm{\mu m}$ distance between each arm and antenna, respectively. (d)  Corresponding PFM imaging of the ferroelectric state after one domain writing cycle of a down, up and periodic configuration ($\mathbf{P_{n,m}}$, n: writing cycle, m: domain configuration) }
\label{Fig1:ExpSetup}
\end{figure}
The device layout and experimental setup for the creation of a reconfigurable magnonic crystal with our multiferroic/ferromagnetic BFO/LSMO heterostructure are shown in \figref{Fig1:ExpSetup}. In order to maximize the influence of the BFO interfacial polarization state we minimized the thickness of the LSMO layer to $13\,\mathrm{nm}$ where the growth conditions are optimized such that the dynamical magnetic properties are improved. For instance,  we obtained a value of $\propto 6\cdot 10^{-3}$ for the Gilbert damping parameter which is comparable to that of thicker films \cite{Madon2018}. 
Further details on growth and the nanofabrication process are given in the methods section and section 1 of the supporting information (cf. Fig. S1).  
We excite and detect spin waves in the Damon-Eshbach configuration \cite{Damon1961} by means of propagative spin wave spectroscopy \cite{Vlaminck2010} (cf. \figref{Fig1:ExpSetup} (a)) and deduce the transmission inductance $L_{21}$ parameter from the recorded data (cf. Supporting information). In order to both write and read the ferroelectric domain configuration of BFO, we employ piezo-force microscopy (PFM) to define the out-of-plane polarization state between the microwave antennae within an area of $\approx 2\times 2\,\mathrm{\mu m}$  (cf. \figref{Fig1:ExpSetup} (d)). The number of the writing steps starting from the virgin state is labelled as $\mathbf{P_{n,m}}$, where $n$ denotes the writing cycle defined as the sequence of virgin, up, periodic domain polarization states and m the specific configuration in that writing cycle. The antenna arms exhibit a periodicity of $500\,\mathrm{nm}$. Hence, the antennae predominantly excite spin waves 
around $K\approx 6 \,\mathrm{rad/\mu m}$ corresponding to a maximum of the antenna's response spectrum (cf. \figref{Fig1:ExpSetup} (b)). 
Therefore, the present study focuses on a periodic ferroelectric domain pattern with  the  same periodicity of $500\,\mathrm{nm}$. The resulting Bragg wavevector of the periodic ferroelectric pattern then corresponds to spin waves of the wave vector of $6\,\mathrm{rad/\mu m}$, where the down and up domains have a width of $300\,\mathrm{nm}$ and $200\,\mathrm{nm}$, respectively. Matching the periods of the antennae and the ferroelectric domains allows to set the bandgap at the most efficiently excited wavevector.   
Consequently, such a matching optimizes the coupling between the antennae and supports the unambiguous interpretation of the results.
{\color{black} Please note, that at room temperature spin wave excitations in BFO occur  in the sub-THz \cite{Bialek2018} band and cannot be coherently excited here. Thus, any observation of a spin wave transmission signal can be fully attributed to the spin wave excitation in the LSMO layer.}
\\
\begin{figure}
\centering
\includegraphics[scale=0.23]{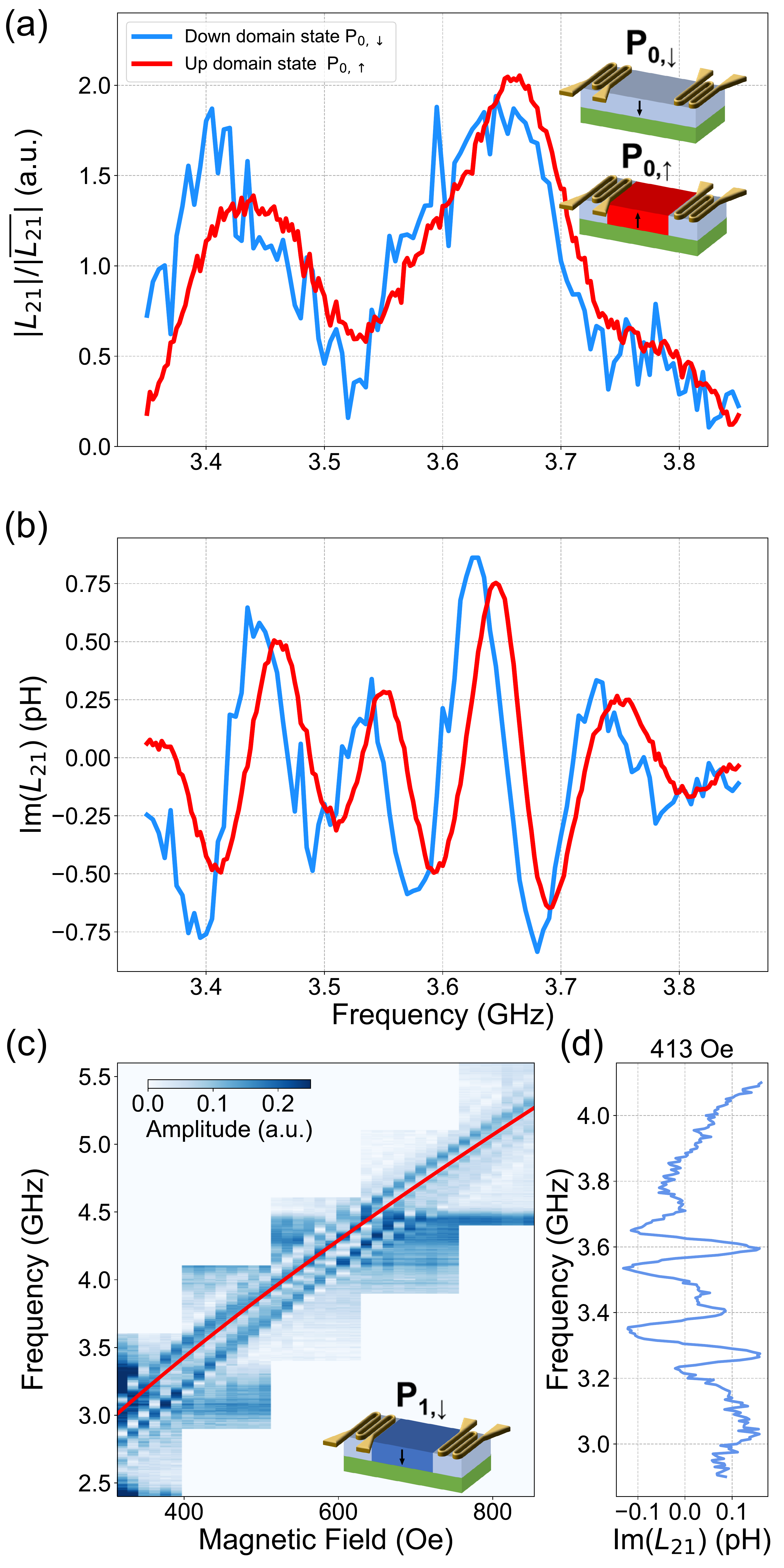}
\caption{(a) Spin wave transmission measurement showing the inductance amplitude $|L_{21}|$ normalized by the mean value for the uniform domain states before writing the periodic state and subsequent reconfiguration. The insets show the corresponding domain state.  (b) Imaginary part of the spin wave signal of $L_{21}$. In (a) and (b) the $\mathbf{P_{0,\downarrow}}$ (blue) state is compared to the uniform $\mathbf{P_{0,\uparrow}}$ (red) state. Apart from a small frequency shift ($\approx 15\,\mathrm{MHz}$) and small change in the amplitude the spin wave propagation is reversible between the uniform domain states of the BFO. (c) Colormap of $(|L_{21}|)(H_{\mathrm{ext}},\omega)$ for the reconfigured $\mathbf{P_{1,\downarrow}}$ state after switching from the up state to the periodic configuration {\color{black} and back to the uniform down state}. The red curve is the theoretical $\omega(H_{\mathrm{ext}},k_x={K}_B)$(cf. equation. \eqref{DE_Dispersion}). (d) $\mathrm{Im}(L_{21})$ of the $\mathbf{P_{1,\downarrow}}$ state at $413\,\mathrm{Oe}$. 
All measurements were conducted at $T=300\,\mathrm{K}$ between ${3.3-3.9\,\mathrm{GHz}}$.  The field independent background is attributed to residual radio frequency standing waves after a VNA short-open-load-through calibration.  }
\label{Fig.2SpinWavePropagationLSMO}
\end{figure}
\figref{Fig.2SpinWavePropagationLSMO} (a) and (b) shows the $L_{21}$ inductance signal of the spin wave for the two uniform up (red) and down polarized (blue) ferroelectric domain states in the BFO layer for the amplitude (in linear scale) and the imaginary part, respectively. The initial (virgin) down state is compared to the up state. 
Apart from small shifts in resonance frequency, 
there is no change between both uniform up and down ferroelectric domain states in the BFO indicating that the PFM writing process does not affect the LSMO properties. We attribute the small frequency difference to a slight misalignment of the sample with respect to the magnetic field between different PFM writing cycles.
Furthermore, since we used less averaging cycles for the S-parameters during the measurement, the noise floor of the virgin state is higher than for the uniform up (cf. \figref{Fig.2SpinWavePropagationLSMO} \textbf{b}) state. Hence, by the comparison of the two different uniform {\color{black}ferroelectric} polarization states before writing a periodic structure (cf. \figref{Fig.2SpinWavePropagationLSMO} \textbf{a-b}) and after rewriting back to a uniform state (cf. \figref{Fig.2SpinWavePropagationLSMO} \textbf{c-d}), we can reliably and reproducibly switch between the various ferroelectric polarization domain states $\mathbf{P_{\mathrm{n,m}}}$ which are also temporally stable.
\newline
In order to model the dispersion relation (cf. \figref{Fig.2SpinWavePropagationLSMO} (c), red line) for our system, we utilize the approach developed {\color{black} by B.A. Kalinikos and A. N. Slavin} \cite{Kalinikos1986,Kalinikos1990} by considering both an in-plane (ip) and out-of-plane (oop) {\color{black}(uniaxial)} anisotropy for spin waves in the Damon-Eshbach configuration which reads as: 
\begin{align}
\omega_n^2=(\Omega_n+\omega_M(1-P_{\mathrm{nn}})+\omega_M\cdot \nonumber \\
 N_{zz})\cdot(\Omega_n+\omega_M\cdot P_{\mathrm{nn}}\cdot \nonumber \\ 
\sin^2(\phi-\phi_{\mathrm{n}})+\omega_M\cdot N_{xx})  \label{DE_Dispersion}
\end{align}
where  $\Omega_{\mathrm{n}}=\omega_{H_n}+\omega_{H_u^{\mathrm{ip}}} \cdot \cos^2(\phi-\phi_0)$ with $\omega_{H_u^{\mathrm{ip}}}=\gamma H_u$ for the {\color{black} {in-plane (ip) anisotropy field}
} and $\omega_{H_n}=\gamma((H_{\mathrm{ext}}+H_{\mathrm{demag}})+M_{\mathrm{s}} \cdot L_{\mathrm{ex}}^2 \cdot K_n^2)$ ($H_{\mathrm{ext}}$: external field, $H_{\mathrm{demag}}$: demagnetization field, $L_{\mathrm{ex}}$: exchange length, $K_n=\sqrt{k_x^2+k_n^2}$, wave vector of spin wave), $\omega_M=\gamma M_{\mathrm{s}}$, $\phi_n=\arctan \frac{k_n}{k_x}$ ($k_n=\frac{n\pi}{w}$, w: width of waveguide), $\phi$ the in-plane angle between the saturation magnetization (here along the y-axis) and the waveguide axis (x-axis), N$_{zz}=-H_u^{\mathrm{oop}}/M_s${\color{black}, where $H_u^{\mathrm{oop}}$ denotes the out-of-plane anisotropy field}, {\color{black} N$_{xx}=-H_u^{\mathrm{ip}}/M_s\sin^2(\phi-\phi_0)$} and the matrix elements $P_{\mathrm{nn}}=1-(\frac{1}{K_n t}(1-e^{-K_n t}))$ (t: ferromagnetic layer thickness) \cite{Kalinikos1986}. 
The modeled dispersion is shown (red solid line) ($H_{\mathrm{demag}}=-13.7\,\mathrm{Oe}$, $k_{\mathrm{x}}=6 \,\mathrm{rad/\mu m}$, $L_{\mathrm{ex}}=5.89\,\mathrm{nm}$, {\color{black}$H_u^{\mathrm{ip}}=100\,\mathrm{Oe}$,  $H_u^{\mathrm{oop}}=-100\,\mathrm{Oe}$, $H_{\mathrm{offset}}=-100\,\mathrm{Oe}$ an offset taking additional contributions to the effective field and uncertainties in the determination of the saturation magnetization  into account}) together with the recorded dependence of $\omega_{n=1}(H)$ for the lowest mode number $n=1$ and $\mathrm{Im \,(L}_{21})$ at $413\,\mathrm{Oe}$ in \figref{Fig.2SpinWavePropagationLSMO} \textbf{c-d} for the $P_{1,\downarrow}$-BFO domain state. 
By employing ferromagnetic resonance, we find $4{\mathrm{\pi}} M_{\mathrm{eff}}=(3323\pm\mathrm{23} )\,\mathrm{G}$ and a Gilbert damping $\alpha_{\mathrm{G}}$ as low as $6\cdot 10^{-3}$ of our 13 nm thick LSMO layer (Supporting information section 2 (cf. Fig. S2)). 
Furthermore, the low value of $\alpha_{\mathrm{G}}$ allows us to extend the distance between the emitting and receiving antennae for the spin wave transmission signal to the micrometre range. In a similar fashion to Ref. \cite{Vlaminck2010}  we can both theoretically and experimentally estimate the spin wave attenuation length of our system. Theoretically, for 
$k_x\approx 6\,\mathrm{rad/\mu m}$, group velocity $v_g\approx 350 \,\mathrm{m/s}$ estimated from our dispersion relation and $\omega/2\pi=3.54\,\mathrm{GHz}$, we obtain $L_{\mathrm{att}}^{\mathrm{th}}=v_g \tau \approx {\color{black}2.1} \,\mathrm{\mu m}$, where $\tau$ is the lifetime defined as $1/\tau = \alpha_{G} (\omega_{H_n} +\frac{\omega_M}{2}(1+N_{xx}+N_{zz}))$. We experimentally estimate the attenuation length from the decay of the spin wave amplitude, \textit{i.e.} $2|\Delta L_{21}|/\Delta L_{11}=e^{(-D_{\mathrm{eff}}/L_{\mathrm{att}})}$, where $D_{eff}=D+D_0$ corresponds to the effective distance between the antennae, with the antenna spacing denoted as $D$ and as $D_0$ the width of the antennae arms, respectively. Then, we obtain $L_{\mathrm{att}}^{\mathrm{exp}}\approx 2 \,\mathrm{\mu m}$  very close to the theoretically expected value which validates the overall modeling approach. 
\\
As a result, the study of the spin wave dispersion spectrum allows now to directly attribute changes in the spin wave transmission signal for periodically altered ferroelectric domains to a periodic modulation, \textit{i.e.} the emergence of a bandgap (cf.  \figref{Fig3GapIndication}). 
\begin{figure}[t!]
\centering
\includegraphics[scale=0.35]{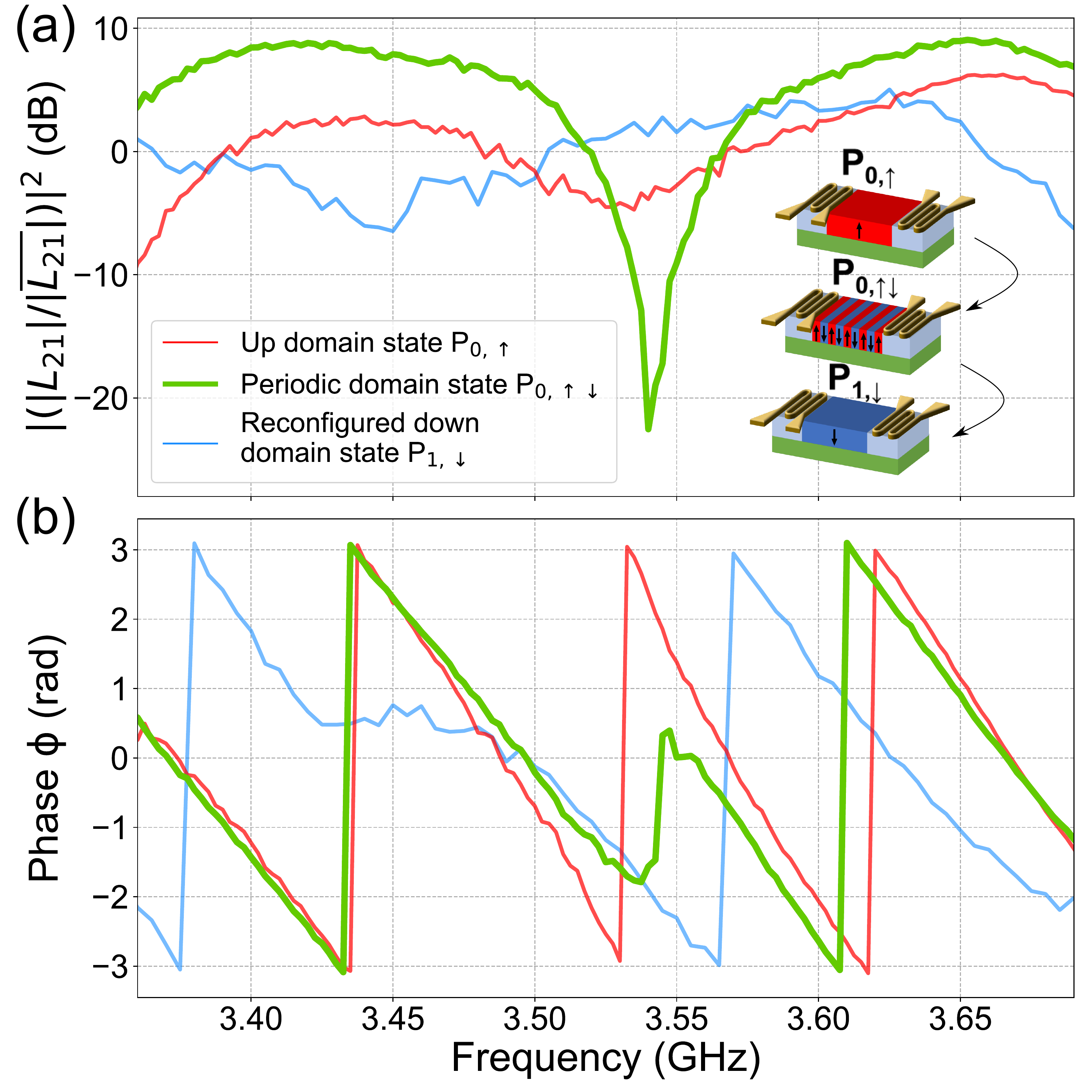}
\caption{\textbf{} (a) Amplitude representation of the reconfigurable magnonic crystal in the BFO/LSMO heterostructure. The transmitted spin wave amplitudes $|\mathrm{L}_{21}|$ are shown in logarithmic scale for the uniform $\mathbf{P_{0,\uparrow}}$ (red), periodic $\mathbf{P_{0,\uparrow \downarrow}}$ (green) and uniform $\mathbf{P_{1,\downarrow}}$ (blue) state  (cf. inset) and compared. The amplitudes are normalized by their mean value. The frequency filtering from the magnonic crystal emerges at $\approx 3.54\,\mathrm{GHz}$ where the spin wave amplitude drops by $20\,\mathrm{dB}$ for the periodic $\mathbf{P_{0,\uparrow \downarrow}}$ state compared to the uniform $\mathbf{P_{0,\uparrow}}$ state. The gap vanishes for reversing the FE domain state back to a uniform down state.
(b) Display of the wrapped phase data for the uniform $\mathbf{P_{0,\uparrow}}$, periodic $\mathbf{P_{0,\uparrow \downarrow}}$ and uniform $\mathbf{P_{1,\downarrow}}$ state. 
 The magnonic bandgap corresponds to the additional phase jump at the bandgap frequency which vanishes again after resetting the FE domain to uniform down ($\mathbf{P_{1,\downarrow}}$). The data is shown for a magnetic field of  ${415} \,\mathrm{Oe} $. Measurements are conducted at $300 \,\mathrm{K}$.}
\label{Fig3GapIndication}
\end{figure}
In \figref{Fig3GapIndication} \textbf{(a)-(b)}, we show the transmitted spin wave's envelope (amplitude, (a)) and phase ((b)) in the LSMO layer from the measurement for the periodic arrangement of up and down polarized ferroelectric domains in the BFO layer {\color{black}(green)} and compare it both to the uniformly up polarized state (red) and reconfigured down polarized state (blue).  
At $\omega/2\pi=3.54\,\mathrm{GHz}$ (cf. \figref{Fig3GapIndication} (a)), the spin wave amplitude drops by $\approx 20\,\mathrm{dB}$ compared to the amplitude in the uniformly up polarized case. Further, at that frequency, we also observe an additional phase jump in the phase spectrum for the case of the periodic domain pattern. The occurrence of such an additional phase jump was not observed for the uniform ferroelectric domains of the BFO layer, where both amplitude and phase show no discontinuity over the measured frequency window. In particular, the gap and the corresponding phase discontinuity due to the change in group velocity vanish by the following reconfiguration of the system back to the uniform $\mathbf{P_{1,\downarrow}}$ FE polarization state. 
Furthermore, both the frequency of the gap and the phase jump are field dependent. As illustrated in \figref{Fig4:GapFieldDependence} we observe a shift to higher frequencies according to the dispersion relation, when the external magnetic field is increased, confirming the bandgap with a strong amplitude rejection exceeding $20\,\mathrm{dB}$.  
\begin{figure}[t]
    \centering
    \includegraphics[scale=0.35]{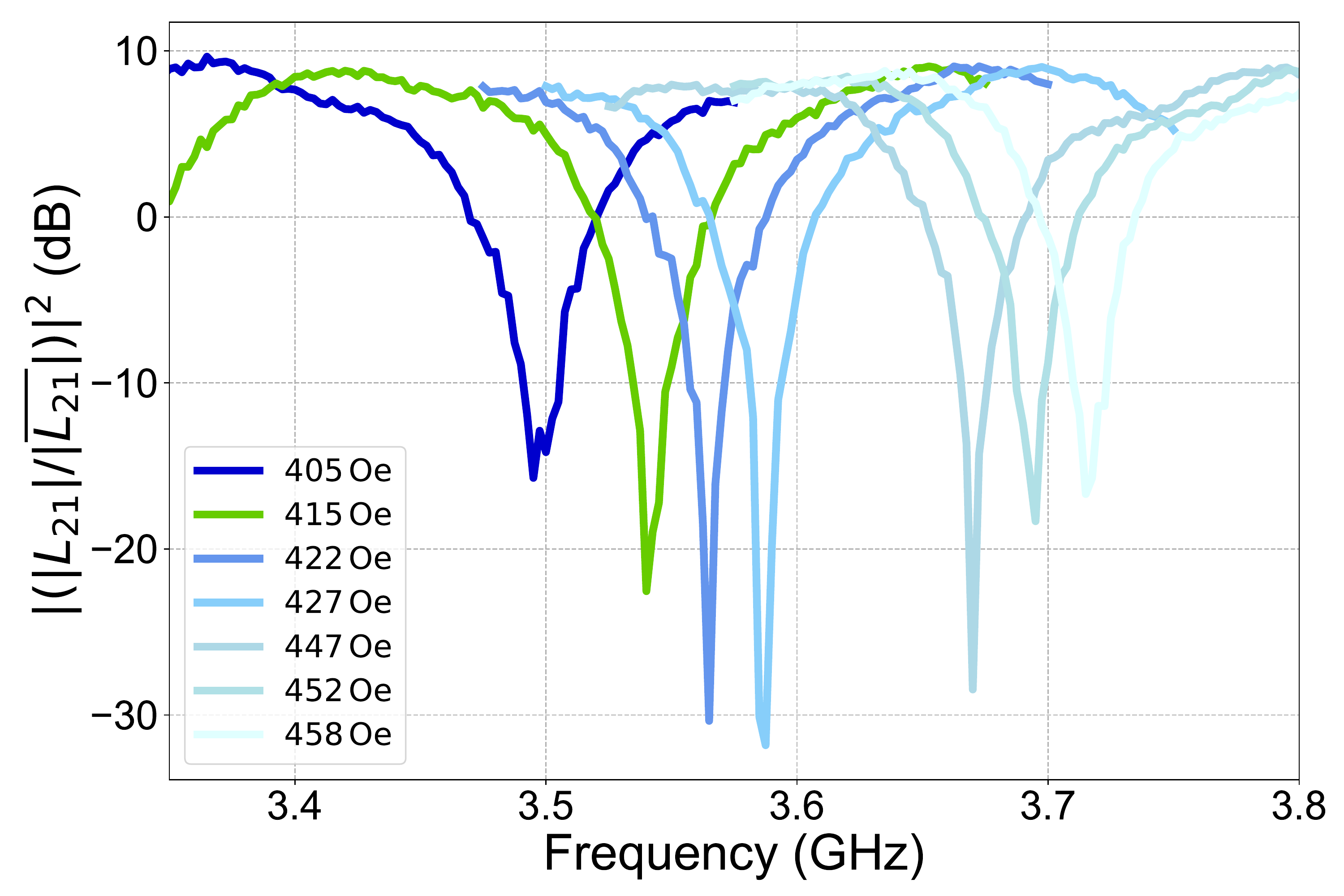}
    \caption{Field dependence of the magnonic bandgap shown for magnetic field values from $405-458 \,\mathrm{Oe})$. The total amplitude varies due to the remaining, field independent background in the system. Hence, the amplitude of the frequency  filtering  changes. This variation is also present in the uniform domain state}. 
    \label{Fig4:GapFieldDependence}
\end{figure} 
\\
The propagative spin wave spectroscopy used here does not provide direct access to the wavevector of the excited spin wave. However, it is possible to infer the wavevector through the phase $\phi$ of the dynamical magnetization $m$.  It depends on the wavevector \textit{\textit{via}} $m\propto e^{i\omega t-k_x(\omega)\cdot x}$, \textit{i.e.} is described as a plane wave, which allows us to assess discontinuities in the dispersion of a MAC. The distance from the antenna exciting the spin wave which propagates to the second, receiving antenna is $D_{\mathrm{eff}}$. As a result, the accumulated phase between $x=0$ and $x=D_{\mathrm{eff}}$ corresponds to $\phi (\omega)={-k_x (\omega)  D_{\mathrm{eff}}}$. Then, the frequency dependence of wavevector can be derived from the unwrapped phase $\phi$ (typically the VNA measures the wrapped phase) by $k_x(\omega)=-\phi/D_{\mathrm{eff}}+K_0$, where $K_0$ is an integration constant that is set by the phase of the excitation field. 

However, by introducing a spacial periodicity, we expect the dispersion relation to split in different branches separated by a magnonic band gap. 
Interestingly in the vicinity of $k_x={K_B}$ (the MAC's Bragg wavevector) the phase relation established previously no longer holds due the branches degeneracy. Indeed, the reflections induced by the periodic domains strongly interfere near the band gap (Bragg conditions). Standing spin-waves are formed, thus the plane wave ansatz does not hold anymore.  Additionally, the position of those standing waves can abruptly change the coupling with the antennas and the value $D_{\mathrm{eff}}$ when they move from the lower to the upper dispersion branch. Hence, the step-like phase jump for the MAC ($\mathbf{P_{0,\uparrow \downarrow}}$ state) is reminiscent of an additional large phase accumulation, probably much larger than $2\pi$ as the 20dB drop in the transmitted amplitude indicates. Measuring the value of this phase-jump would require a time-resolved measurement scheme.
Our observations demonstrate that a change in the ferroelectric state unquestionably modulates the spin wave dispersion of the ferromagnetic layer. \\
However, the underlying governing physical mechanism that couples the polarisation state of BFO and LSMO is still to be identified. 
The most probable mechanism would imply the coupling of the ferroelectric state to the antiferromagnetic order in BFO that induces a modulation of an the interfacial exchange field. As a matter of fact BFO/LSMO heterostructures are known to exhibit magneto-electric control of the exchange bias field (up to $160\,\mathrm{Oe}$ at $10\,\mathrm{K}$) but this field vanishes when the temperature is increased on macroscopic devices towards room temperature \cite{Wu2010, Yi2019}. However, given the small sizes of the written ferroelectric domains ($200\,\mathrm{nm}$ and $300\,\mathrm{nm}$) in our system, it is possible that there is a remanent exchange coupling between BFO and LSMO that is sufficient in amplitude to generate a MAC and enables the observed frequency filtering. 
Correspondingly, as well as a thickness dependent study of the LSMO layer in addition to the modulation of the  ferroelectric polarization of the BFO and a temperature dependent study on the role of the exchange bias could give further insight into the relevance of these effects and the role of the interface for these types of oxide heterostructures. 
\section*{Conclusion}
In summary, we implemented a tuneable MAC filtering scheme based on BFO/LSMO thin film heterostructures. Our approach allows us to reliably and reversibly filter certain frequencies in the spin wave dispersion, analogous to magnonic crystals, by  controlling the ferroelectric domain state order of the BFO layer. Once written in a certain domain pattern., no further energy input is required in the MAC filtering, which results in minimal power consumption for the operation of our device. 
Furthermore, the writing voltage does not exceed $5\,\mathrm{V}$. 
Thus, the introduction of multiferroics, reconfigurable and low power voltage controlled materials into microscopic magnonic systems represents a key step towards large scale integrated magnonic systems.

\section*{Methods}
\subsection*{Growth}
Both LSMO and BFO are perovskites with well-matched unit cell parameters. They are epitaxially grown on an NGO substrate using pulsed laser deposition (PLD) \cite{Bea2006}. 
Please note, that in order to reduce leakage effects of the film between the electrode and the PFM tip, the BFO layer is not pure BFO but has been doped with $5\,\%$ of Manganese (Mn).
Regarding the growth conditions, the BFO/ LSMO //NGO (001) is grown by pulsed laser ablation under oxygen-rich growth conditions of $800\,\mathrm{^{\circ}C}$ and $0.2\,\mathrm{mbar}$ for the LSMO film, and $650\,\mathrm{^{\circ}C}$ and $6\cdot 10^{-3}\,\mathrm{mbar}$ for the BFMO film. The frequency of the laser is $2\,\mathrm{Hz}$ and $5\,\mathrm{Hz}$, respectively. Annealing at $650\,\mathrm{^{\circ}C}$ is carried after the growth of the bilayer for $30$ minutes in an environment rich in oxygenates ($200\,\mathrm{mbar}$).
\subsection*{Device fabrication}
For the fabrication of our device we employed a combination of laser lithography and electron ion beam machining for the smallest structures such as the microwave antennae. As for the waveguide's dimensions it exhibits a width of $2\, \mu m$ and length of $90\,\mu m$. 
A thin ($d_{\mathrm{AlO_3}}=10\,\mathrm{nm}$) alumina oxide ($\mathrm{AlO}_3$) layer is sandwiched between the BFO and a titanium ($d_{\mathrm{Ti}}=20\,\mathrm{nm}$) layer. 
The titanium layer enhances the adhesion of the gold for the addition of the microwave antennae for the propagative spin wave spectroscopy. These antennae (Au, $d=130\,\mathrm{nm}$) exhibit a meander-like structure with a periodicity of $500\,\mathrm{nm}$ with several periods narrowing the wavevector band of excited spin waves.  
\subsection*{Propagative spin wave spectroscopy}
The spin wave spectra are obtained \textit{via} probe-tip ferromagnetic resonance. The resulting transmission and reflection scattering matrix $S_{ij}$ is recorded by means of network analysis employing a two-port (i and j) vector network analyzer setup.
{\color{black} \section*{Supporting information}
We have provided supporting information for our work which is available online.
The contents are:
\begin{itemize}
\item Material’s characterizations including a RHEED diffraction pattern. 
\item Information on the measured scattering parameters and deduced inductance matrix
\item Characterization of dynamic magnetic properties: Ferromagnetic resonance measurements of the unpatterned BFO/LSMO bilayer.
\end{itemize}
 }
\section*{Acknowledgements}
This work has been supported from the European  Union’s  Horizon  2020  research  and  innovation  program within the FET-OPEN project CHIRON under grant agreement No.  801055. VH is grateful for support from the project OISO under grant agreement ANR-17-CE24-0026 from the Agènce National de la Recherche.
\section*{Author contributions}
V.H and C.C. grew and characterized the sample with H.M. H.M, L.V.patterned the devices. 
I.B and H.M characterized the devices and conducted the measurements. I.B, H.M, M.A  analysed the data and wrote the manuscript. 
A.V., S. F and V.G. performed the PFM characterization and writing. M.A., A.B., P.B. and M.B. supervised the project. All authors discussed the results and participated in the preparation of the manuscript.


%% file: SupportingInformation_revised_cleancopy_arxiv.tex
\title {{\color{black} Supporting Information: }Voltage-Controlled Reconfigurable Magnonic crystal at the Submicron Scale}

\author{Hugo Merbouche}
\altaffiliation{These authors contributed equally}
\affiliation{Unit\'{e} Mixte de Physique CNRS, Thales,  Universit\'{e}  Paris-Saclay, 91767 Palaiseau, France}

\author
{Isabella Boventer$^{*}$}

\email{Corresponding author: isabella.boventer@cnrs-thales.fr}

\affiliation{Unit\'{e} Mixte de Physique CNRS, Thales,  Universit\'{e}  Paris-Saclay, 91767 Palaiseau, France}

\author{Victor Haspot}

\altaffiliation{These authors contributed equally}
\affiliation{Unit\'{e} Mixte de Physique CNRS, Thales,  Universit\'{e}  Paris-Saclay, 91767 Palaiseau, France}


\author{Stéphane Fusil}
\affiliation{Unit\'{e} Mixte de Physique CNRS, Thales,  Universit\'{e}  Paris-Saclay, 91767 Palaiseau, France}

\address{Universit\'{e} d’Evry, Universit\'{e} Paris-Saclay, 91000 Evry, France}
\author{Vincent Garcia}
\affiliation{Unit\'{e} Mixte de Physique CNRS, Thales,  Universit\'{e}  Paris-Saclay, 91767 Palaiseau, France}

\author{Diane Gou\'{e}r\'{e}}
\affiliation{Unit\'{e} Mixte de Physique CNRS, Thales,  Universit\'{e}  Paris-Saclay, 91767 Palaiseau, France}

\author{Cécile Carr\'{e}t\'{e}ro}
\affiliation{Unit\'{e} Mixte de Physique CNRS, Thales,  Universit\'{e}  Paris-Saclay, 91767 Palaiseau, France}

\author{Aymeric Vecchiola}
\affiliation{Unit\'{e} Mixte de Physique CNRS, Thales,  Universit\'{e}  Paris-Saclay, 91767 Palaiseau, France}

\author{Romain Lebrun}
\affiliation{Unit\'{e} Mixte de Physique CNRS, Thales,  Universit\'{e}  Paris-Saclay, 91767 Palaiseau, France}

\author{Paolo Bortolotti}
\affiliation{Unit\'{e} Mixte de Physique CNRS, Thales,  Universit\'{e}  Paris-Saclay, 91767 Palaiseau, France}

\author{Laurent Vila}

\address{Universit\'{e} Grenoble Alpes, CEA, CNRS, Grenoble INP, Spintec, 38000 Grenoble, France}
\author{Manuel Bibes}
\affiliation{Unit\'{e} Mixte de Physique CNRS, Thales,  Universit\'{e}  Paris-Saclay, 91767 Palaiseau, France}

\author{Agnès Barth\'{e}l\'{e}my}
\affiliation{Unit\'{e} Mixte de Physique CNRS, Thales,  Universit\'{e}  Paris-Saclay, 91767 Palaiseau, France}

\author{Abdelmadjid Anane}
\email{Corresponding author: madjid.anane@universite-paris-saclay.fr}
\affiliation{Unit\'{e} Mixte de Physique CNRS, Thales,  Universit\'{e}  Paris-Saclay, 91767 Palaiseau, France}
\begin{abstract}
In this {\color{black}supporting information}, we give more detailed information on the material's characterization both in terms of the surface quality and the dynamic magnetization properties \textit{via} ferromagnetic resonance measurements.
\end{abstract}
\maketitle
\section*{Methods}
\subsection*{BFO/LSMO film characterization}
\begin{figure}[h!]
\centering
\includegraphics[scale=0.5]{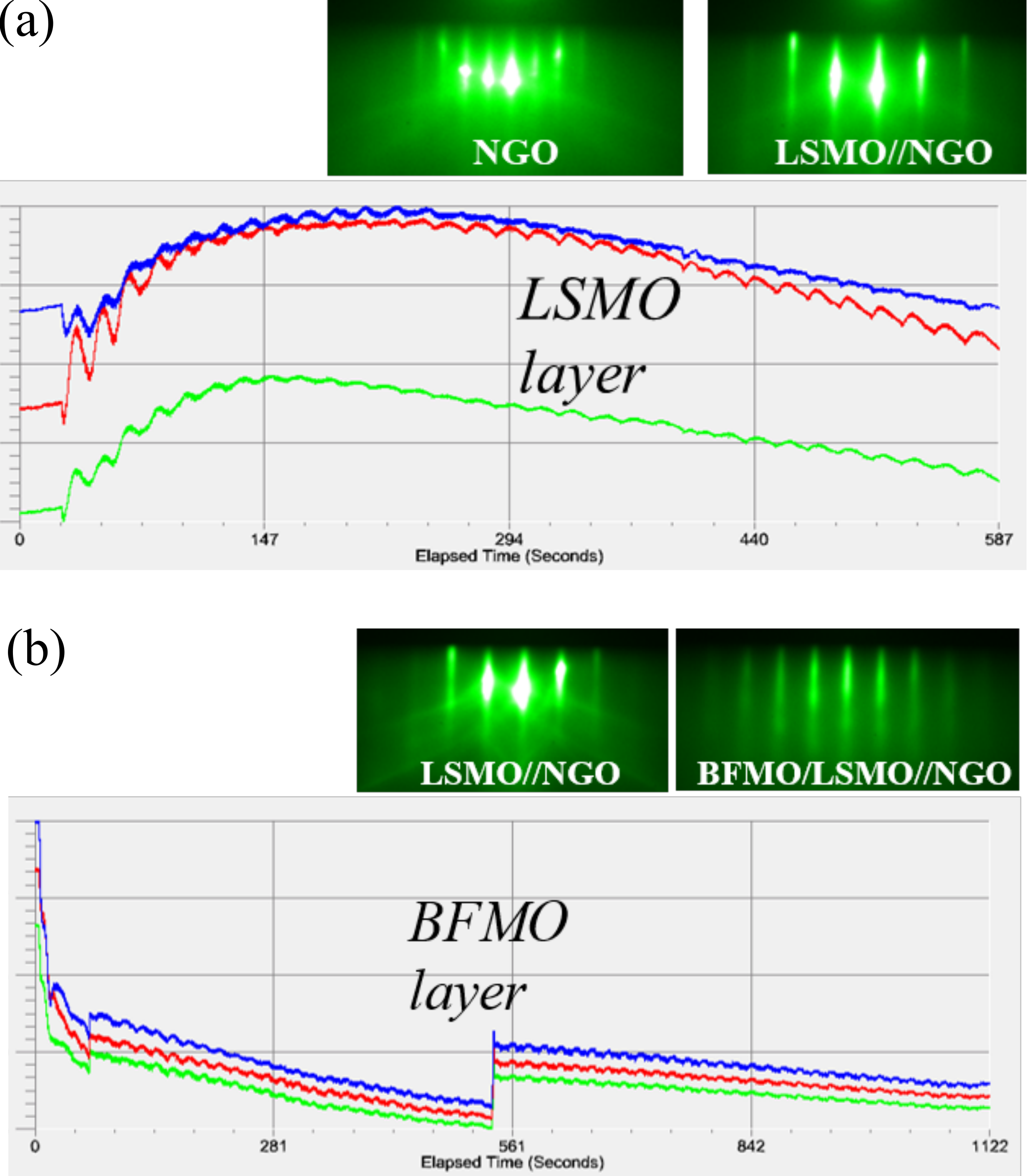}
\caption{
(a) Clear Rheed pattern of solely LSMO on NGO and (b) Rheed pattern after the addition of the BFO layer }
\label{S1}
\end{figure}

 In order to characterize the surface quality of both components of our heterostructure, we performed reflection high-energy electron diffraction (RHEED) spectroscopy. The spectroscopic data is shown in \figref{S1} for LSMO on the NGO substrate only ((a)) and for the complete heterostructure with BFO ((b)). 
Both the RHEED patterns are well separated from each other, with an elongation in the vertical direction indicating a flat surfaces.  
Please note that, the uniaxial in-plane anistropy $H_u^{\mathrm{ip}}$ in the LSMO originates from our orthorhombic NGO substrate with a (001)/K(110) orientation which induces an anisotropic strain, where K denotes the anisotropy  plane. For this substrate, the LSMO grows with an in-plane angle of $45^{\circ}$ to the [100] orthorombic direction in NGO in (001) pseudocubic configuration.\cite{Boschker2009}  
As a result, the BFO/LSMO waveguides exhibit a [110] uniaxial anistropy with respect to the LSMO crystallographic direction, yielding an angle $\phi_0$  of $45^{\circ}$ between the direction of the easy axis and the waveguide axis (x-axis) which needs to be included in the dispersion relation. 
In addition, we include an out-of-plane anisotropy $H_u^{\mathrm{oop}}$ (along the z-axis) due to the magneto-elastic anisotropy induced by the slight lattice mismatch between LSMO and the NGO substrate leading to a compressive strain for our thin LSMO film. \cite{Xiao2019} The magnetic field is along the y-axis direction, in the DE configuration.

\section*{Measured and deduced parameters}
Specifically, we measure the device's properties by probe-tip ferromagnetic resonance (FMR) with a vector network analyser (VNA) by recording the scattering matrix S for a two-port system. \cite{Kalarickal2006}
Afterwards, for the discussion of spin wave propagation and modulation in LSMO, similar to Vlaminck \textit{et al.}, \cite{Vlaminck2010_2} we calculate the corresponding inductance matrix L from our raw data as:  
\begin{eqnarray}
\mathrm{L}=
\begin{pmatrix}
\mathrm{L}_{11}& \mathrm{L}_{12} \\
\mathrm{L}_{21} & \mathrm{L}_{22}
\end{pmatrix}\\
=\frac{\mathrm{Z}_0}{\delta}
\begin{pmatrix}
\alpha+\mathrm{S}_{21}\cdot \mathrm{S}_{12} & 2 \mathrm{S}_{21}\\
2\mathrm{S}_{12} &\beta +\mathrm{S}_{21}\cdot \mathrm{S}_{12}
\end{pmatrix}
\end{eqnarray}, where $\delta=\omega_0[(1-\mathrm{S}_{11})(1-\mathrm{S}_{22})-\mathrm{S}_{12}\cdot \mathrm{S}_{21}]$ with the resonance frequency $\omega_0$, $\alpha=(1+\mathrm{S}_{11})(1-\mathrm{S}_{22})$, $\beta=(1+\mathrm{S}_{22})(1-\mathrm{S}_{11})$ and $\mathrm{Z}_{0}$ is the impedance.
\section*{Ferromagnetic resonance study on utilized heterostructure}
\begin{figure}[h!]
\centering
\includegraphics[scale=0.52]{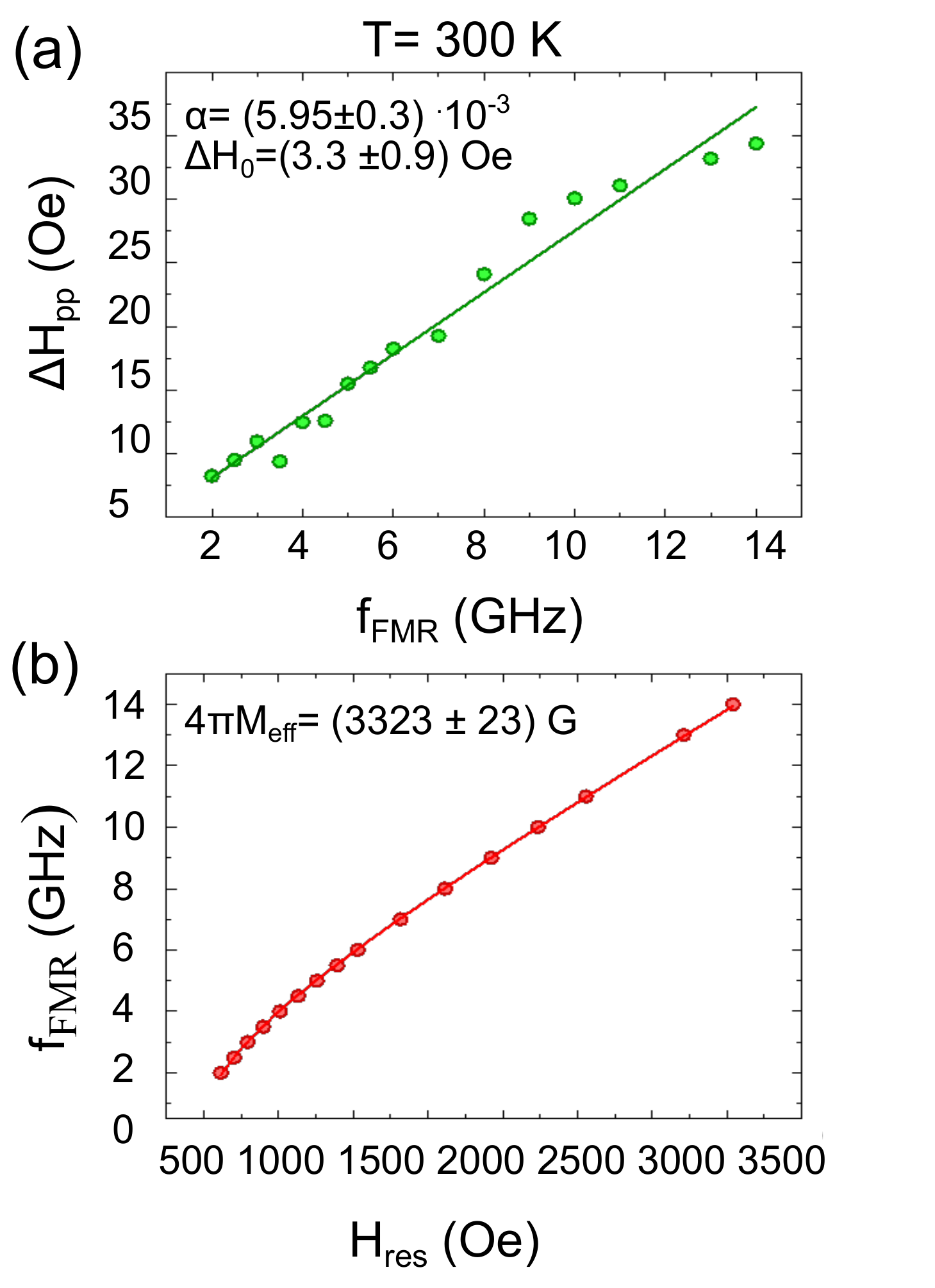}
\caption{Magnetic characterization of the unpatterned BFO/LSMO heterostructure  measured with field-modulated ferromagnetic resonance at room temperature.
(a) Frequency dependence of the linewidth for the  
determination of the Gilbert damping parameter ($\alpha_{\mathrm{G}}=(5.95\pm 0.3)\cdot 10^{-3}$) and the intrinsic linewidth at zero applied external field of the specific film ($\Delta H_0=(3.3\pm 0.9)\,\mathrm{Oe}$). 
(b) Ferromagnetic resonance frequency as a function of the applied static magnetic field and determination of the effective  magnetization ($4\pi M_{\mathrm{eff}}=(3323\pm 23)\,\mathrm{G}$). }
\label{S2}
\end{figure}
In order to determine the saturation magnetization, the Gilbert damping parameter and the contribution to the linewidth from inhomoegenous broading denoted by $\Delta H_0$, we performed ferromagnetic resonance (FMR) measurements on our pure films before any patterning. In \figref{S2} (a), we show the frequency dependence of the peak-to-peak linewidth $\Delta H_{\mathrm{pp}}$ including a linear fit to determine the damping. As mentioned in the main text, the effective Gilbert damping of our heterostructure made of a BFO/LSMO//NGO stack is determined to $\alpha_{\mathrm{G}}\approx 6\cdot 10^{-3}$. 
In addition, the dispersion at $T=300\,\mathrm{K}$, \textit{i.e.} $\omega_{\mathrm{FMR}}(H_{\mathrm{res}})$ including a fit of the Kittel law for thin films is shown in \figref{S2} (b). The fit yields $4\pi M_{\mathrm{eff}}= (3323\pm 23) \,\mathrm{G}$. A comparison of that value to the Probe-Tip FMR measurements of the frequency dispersion of the transmitted spin wave and using the analytical formula (cf. Equation (3) in main part) shows an agreement between the experimental and analytically derived numbers.     
Please note, that due to the required \textit{in situ} growth of LSMO and subsequently of BFO during the pulsed laser deposition process, only the FMR of the complete heterostructure could be measured. Thus, the obtained values are effective values of the heterostructure. 
The Curie Temperature has been also measured by resistivity measurements and was found to be $T_c\approx 390\,\mathrm{K}$.
